\documentstyle{article}
\newcommand {\eq}{\begin{equation}}
\newcommand {\qe}{\end{equation}}

\newcommand {\ea} {{\it et al.}}

\newcommand {\prc}{Phys. Rev. C}
\newcommand {\prd}{Phys. Rev. D}
\newcommand {\h}{\frac{1}{2}}

\newcommand {\pii} {$\pi$}

\newcommand {\pr}{ Phys. Rev. }
\newcommand {\np}{Nucl. Phys. }
\newcommand {\prl}{ Phys. Rev. Lett. }
\newcommand {\pl}{ Phys. Lett. }
\newcommand {\nucp}{Nucl. Phys. }

\newcommand {\qbq}{\bar{q}q}
\begin{document}
\title{The Contribution of the Light Quark Condensate to the $\pi$N Sigma 
Term}

\author{ W. R. Gibbs\\ 
Department of Physics, New Mexico State University \\
 Las Cruces, New Mexico 88003, USA\\gibbs@nmsu.edu\\}

\maketitle

\begin{abstract}

There has been a discrepancy between values of the pion-nucleon sigma term
extracted by two different methods for many years. Analysis of recent high
precision pion-nucleon data has widened the gap between the two
determinations. It is argued that the two extractions correspond to 
different quantities and that the difference between them can be 
understood and calculated.

\end{abstract}

\section{Introduction}

The sigma term is directly related to explicit chiral symmetry breaking in
the nucleon. The conventional view has been that its value can be obtained 
by either an extrapolation of the isospin zero pion-nucleon scattering
amplitude to a defined subthreshold point or by comparing masses of 
members of the baryon octet.  There has long been a problem that the two 
determinations do not result in the same value.  

The value extracted from the comparison of masses ($\Sigma_M$) is commonly
thought to be around 35 MeV while the value of the sigma term extracted from
pion-nucleon scattering ($\Sigma_S$) was given by Koch as $64 \pm 8$ MeV
\cite{koch} using dispersion analysis and the Karlsruhe-Helsinki phase
shifts.  While this difference was disturbing, there were theoretical
corrections to be made and the pion-nucleon data at that time were not of 
high quality, especially at low energy.

It was suggested that the difference could be interpreted as evidence
for a large component of strangeness in the 
nucleon\cite{donoghue,sainio}. Estimates of the strangeness content from 
the strange meson cloud give around 7.6\%\cite{lyub}. From neutrino 
induced reactions the strangeness content has been reported as 6.4 
\%\cite{smith} and 9.9 \%\cite{bazarko} (although corrections to these 
numbers may be significant\cite{boros}), making the 20\% required for this 
explanation of the discrepancy questionable.

The \pii N data base has been improved recently (see Ref. \cite{gak} for
an analysis).  New sigma-term extractions with this data, however, did not
reduce the discrepancy, but increased it.

Kaufmann and Hite\cite{kh} used interior and fixed $t$ dispersion
relations to map out the subthreshold amplitude. They obtained a value of 
$\Sigma_S=88\pm 15 $ MeV. Recently Olsson and Kaufmann\cite{olsson}, 
using Olsson's sum rule, found $\Sigma_S=80-90$ MeV. Pavan\cite{pavan} 
reported on a study using the newest VPI/GWU\cite{gwu} analysis of 
data which found $\Sigma_S=79\pm 7 $ MeV.

Modern estimates give a 15 MeV reduction\cite{bkm,gls,bl,brown} for the 
correction for the finite value of $t$ at the Cheng-Dashen point. It has 
been suggested that, based on PCAC, this correction may be too 
large\cite{sidmike}. Lattice calculations\cite{dong} give a decrease of 
6.6  MeV. The value of the sigma term extracted from pion-nucleon 
scattering would appear to be $85-15=70\pm7$\ MeV, or perhaps somewhat 
greater.

It is the aim of this paper to suggest that the discrepancy discussed
above can be understood if the effect of the light-quark condensate
is taken into account. This condensate creates an energy density 
throughout free space with the common belief that it is zero within
confining regions, such as the interior of a baryon. As the normal
theory is adjusted for the change in the energy zero in free space this
absence of the condensate in the nucleon effectively gives a contribution
to the baryon mass which is unrelated to its structure (other than the
volume).  Since the light quark condensate is proportional to the light
quark mass it will give a contribution to the sigma term which is equal
to its contribution to the mass of the baryon (in particular the nucleon)
from explicit chiral symmetry breaking.  This correction is investigated
from several points of view in what follows.

\section{The Role of the Condensate}

The evaluation of the sigma commutator with the use of an effective
Lagrangian \cite{chengli} shows it to be proportional to the nucleonic
expectation value of $\qbq=\h(\bar{u}u+\bar{d}d)$, where the light quark
mass difference has been neglected (i.e.  $m_u=m_d=m_q$).

In a theory with spontaneous symmetry breaking one must modify the 
commonly used definition
\eq
\Sigma = 2 m_q \int d^3x <\!\!N|\overline{q}q(x)|N\!\!>
\qe
where the state $|N\!\!>$ is interpreted as a system which is localized in
space within a nucleonic volume. Since outside this region
$<N|\overline{q}q(x)|N>\rightarrow <0|\overline{q}q(x)|0>$, a constant, 
the integral diverges. To obtain a finite result one can follow several 
authors \cite{cohen,gammal,magda,ballot} to define
\eq
\Sigma_S = 2 m_q \int d^3x \left[ <\!\!N|\overline{q}q(x)|N\!\!> - 
<\!0|\overline{q}q(x)|0\!> \right].
\qe

To estimate the size of the vacuum correction within the nucleon,  
consider the model of the nucleon as a constant density over a sphere of 
radius R and volume V and rewrite the last equation as
\eq
\Sigma_S=2m_q\int_V\! d^3x <\!\! N|\qbq(x)|N\!\! >-2m_qV<\!0|\qbq|0\!>.
\label{eq2}
\qe

A nucleon averaged value can be defined as
\eq
\overline{<\!N|\qbq|N\!>}=\frac{1}{V}\int_V d^3x <\!N|\qbq(x)|N\!>.
\qe
It is common to assume that the condensate is zero inside the nucleon so 
that only that part which depends on the structure of the baryon remains.

The GOR\cite{gor} relation leads to the expression 
\eq
2m_q<\! 0|\qbq|0\! >=-\mu^2f_{\pi}^2=-21.7\ {\rm MeV/fm^3}
\label{density}
\qe
where $f_{\pi}$ is the pion decay constant (92.4 MeV).

With the definition
\eq
\delta=-2m_qV<\! 0|\qbq|0\! >=V\mu^2f_{\pi}^2,
\qe
one can write Eq. \ref{eq2} as
\eq
\Sigma_S=2m_qV \overline{<\! N|\qbq|N\! >}+\delta. \label{eq7}
\qe

While the second term may seem unfamiliar, on might remember that the
sigma term obtained from the linear sigma model\cite{sigmod} depends
only on the {\em vacuum} expectation value of the sigma field (often
identified with $\qbq$).

\section{Scattering from a Bubble of Perturbative Vacuum}

As a direct illustration of the effect of the vanishing condensate
in the interior of the nucleon on pion-nucleon scattering, consider
a ``bubble'' model for the s-wave $\pi^0$p (i.e. isospin even)  
scattering amplitude in which the pion wave function satisfies a
Klein-Gordon (KG) equation. Outside of the bubble of radius $R$ the
pion has momentum, mass, and energy given by $k$, $\mu$, and
$\omega=\sqrt{k^2+\mu^2}$. Within the bubble, the interaction of the
pion with the components of the structure is represented by a
constant, energy-independent potential, $v$ which will be introduced
as the fourth component of a Lorentz four-vector.  The effect of the
vanishing of the condensate will be modeled by setting the pion's
effective mass to zero in this region which can be done by choosing
an appropriate Lorentz scalar potential $s$. The KG equation inside
the sphere is
\eq
k_0^2 + (\mu-s)^2 =(\omega-v)^2
\qe
where $k_0$ is the wave number in the interior. By taking $s=\mu$
the mass term is eliminated. To evaluate $v$, the requirement is
imposed that the scattering amplitude vanish at threshold (suggested
by the data and chiral symmetry), which requires that the inner and
outer momenta be equal at this point ($k_0=k=0$) and establishes the
value of $v$ as $\mu$ thus leading to
\eq
k_0^2=(\omega-\mu)^2 .
\qe
The scattering amplitude corresponding to the lowest order solution
of the KG equation (good to a few percent over the energy range
considered here)  is given by
\eq
f_0(k)=\frac{1}{3}(k_0^2-k^2)R^3=\frac{2}{3}\mu(\mu-\omega)
R^3\rightarrow \frac{2}{3}\mu^2R^3
\label{result}\qe
where the arrow indicates evaluation at $\omega=0$ (a very good
approximation to its value at the Cheng-Dashen point). Agreement
of this expression with the low energy amplitude\cite{gak} is 
obtained for $R$ around 0.75 fm.

One can consider the two effects (pion mass zero and the potential
$v=\mu$) as generating separate amplitudes (equal at $\omega=0$) obtained
by setting, in turn, $s$ or $v$ to zero. These two amplitudes add
to give the result of Eq. \ref{result} indicating that the vacuum 
subtraction term and the term arising from scattering from the components 
of the nucleon should give equal contributions to the sigma term.  With 
the factor $4\pi f_{\pi}^2$, $\delta$ is reproduced by the amplitude 
arising from the effect of the zero mass of the pion alone.

\section{Extraction of the Sigma Term from the Baryon Masses}

Consider the determination of the sigma term from baryonic mass  
differences. From Gasser\cite{gasser}, to lowest order in the quark mass, 
\eq
M_n=M_0+m_q(P_u+P_d)+m_sP_s
\label{10}
\qe
\eq
M_{\Sigma^0}=M_0+m_q(P_d+P_s)+m_sP_u
\qe
\eq
M_{\Xi^0}=M_0+m_q(P_u+P_s)+m_sP_d
\label{12}
\qe
where $ P_x=V\overline{<\! n|\bar{x}x|n\! >} $
and $|n\!\! >$ is the neutron state vector.

$M_0$ represents the mass of the baryon in the limit of zero quark masses.   
Neglecting the strangeness content of the nucleon and taking the 
appropriate combination of the masses to eliminate $M_0$, Eqs. 
\ref{10}-\ref{12} can be solved for the combination $P_u+P_d$. Applying 
the equation
\eq
\Sigma_{\pi N}=m_q\frac{dM_N}{dm_q} \label{fh}
\qe
to Eq. \ref{10} yields\cite{gasserbis}
\eq
\Sigma_M=m_q(P_u+P_d)=2m_qV\overline{<N|\qbq|N>}=
\label{sigm}
\qe
$$
\frac{m_q}{m_s-m_q}(M_{\Xi^0}+M_{\Sigma^0}-2M_n)=26.3\ {\rm MeV},
$$
which corresponds to the first term in Eq. \ref{eq7}. The numerical value 
results from an assumed quark mass ratio of $m_s/m_q=25$. 
Cheng and Li\cite{chengli} (see Eq. 5.257) obtain a similar result,
\eq
\Sigma_M=\frac{3m_q}{m_q-m_s}(M_{\Lambda}-M_{\Xi^0})=25.9\ {\rm MeV}.
\qe
  
No shift of the masses for the vacuum energy density was considered
in this analysis, but if it were the masses would all be shifted by
the same amount and the shift would not appear in the final
expression. Gasser calculated\cite{gasser} a correction from higher
orders in the quark mass of about 10 MeV (although see Ref.
\cite{jameson}). Thus, the corrected value from the mass analysis is

\eq
\Sigma_M=35 \pm 5\ {\rm MeV}.
\qe

A comparison of Eqs. \ref{eq7} and \ref{sigm} yields
\eq
\Sigma_S=\Sigma_M+\delta.
\qe

To calculate the expected value of $\delta$, the volume of the confinement
region of the nucleon needs to be known.  One estimate of this radius can
be obtained from form factors for pion coupling. For a monopole form
factor, a mass of 800 MeV to 1 GeV was found\cite{coon}.  The volumes
corresponding to these values lead to a vacuum contribution in the range 
$\delta=22.7\ {\rm to}\ 43.2 $\ MeV. From a lattice calculation of the
distribution of valence quarks\cite{lissia} the equivalent square radius
of 0.65 fm can be obtained, leading to $\delta$=25.0 MeV. Using the value
of $R$ found in the ``bubble'' model above $\delta$ is found to be 38 MeV.

\section{Bag Models}

The same physics appears in the bag model where it is supposed that
there is an energy density, B, inside the bag which exerts a pressure
tending to enlarge the bag. B is normally treated as a parameter to be 
fit in the process of matching the model to the data. The exclusion of the 
physical vacuum energy density from the cavity  provides the same effect 
as an energy density within the bag. In theories in which the bag constant 
is calculated from such considerations\cite{gogohiatoki,shuryak1} it can 
be broken into chiral symmetry conserving and breaking parts,
\eq
B=B^0+B^{\chi SB}
\qe

where $B^{\chi SB}$ is given by Eq. \ref{density}. The attempt to
identify the vacuum energy density with the bag constant has a long
history (see e.g. \cite{gogohiatoki,shuryak1,gogohia}). The
difficulty in making a credible identification has been that the
contribution gluons to the vacuum energy density appears to be much
larger than the empirical bag constants (Table I) whereas the energy
density due to the light quark condensate is of the same order.  
Whatever the resolution of this problem, there seems to be general
agreement as to the inclusion of the light quark condensate term in
the vacuum energy density.

\begin{table}
\vspace*{.3in}
\begin{center}
\begin{tabular}{|l|c|}
\hline
Authors& B(MeV/fm$^3)$\\
\hline
\hline
Chodos et al.\cite{chodos1}&27.0\\
\hline
Chodos and Thorn\cite{chodos2}&32.1\\
\hline
Barnhill et al.\cite{barnhill}&39.5\\
\hline
DeGrand et al.\cite{degrand}&57.6, 31.8 (l)\\
\hline
Chanowitz and Sharpe\cite{chan}& 27- 66\\
\hline
Vasconcellos et al.\cite{vasconcellos}& 20 (l)\\
\hline
\end{tabular}
\end{center}
\caption{Bag constants obtained by various groups. The notation (l)
indicates that a large quark mass was used.}
\end{table}
In bag models the mass of the nucleon is given by
\eq
M_N=<N|H_{\rm structure}|N>+BV.
\qe
A number of calculations have been made based on this 
formula\cite{birse,jaffesig,jameson1,lyub} 
where the sigma term was obtained from the first term representing the 
valence and sea quarks. The  contributions from these two component 
scattering terms were found to be roughly equal and add to give a 
contribution to the sigma term of 35 to 45 MeV. Applying Eq. \ref{fh} to 
the second term, as well as the first, the additional $\delta$ 
contribution arises naturally in these models.

Note that spontaneous symmetry breaking (non-zero $<\! 0|\qbq|0\! >$), 
explicit symmetry breaking ($m_q\ne 0$) and a finite size nucleon ($V\ne 0$) 
are all needed for this correction. Any theory which does not include all
three will find zero for $\delta$. Thus, a non-zero value of the condensate
is required and the difference in the two determinations of the sigma term 
might be taken as an indication of its existence.

\section*{Acknowledgments}

I wish to thank W. B. Kaufmann for a great deal of discussion on this 
problem. I also thank B. Loiseau and T. Goldman for helpful discussions 
and T. Cohen for a useful communication.

This work was supported by the National Science Foundation under contract
PHY-0099729.


\begin{thebibliography}{300}

\bibitem{koch} R. Koch, Z. Phys. C15, 161(1982)

\bibitem{donoghue} J. F. Donoghue and C. R. Nappi, \pl 168B, 105(1986)

\bibitem{sainio} M. Sainio, $\pi$N Newsletter {\bf 10}, p. 13(1995)

\bibitem{lyub} V. E. Lyubovitskij, T. Gutsche, A. Faessler and
E. G. Drukarev, \pr D 63, 054026(2001) 

\bibitem{smith} W. H. Smith \ea, \nucp B S31 262(1993)

\bibitem{bazarko} A. O. Bazarko \ea, Z. Phys. C65, 189(1995) 

\bibitem{boros} C. Boros, J. T. Londergan and A. W. Thomas, \prd 58, 
114030(1998)

\bibitem{gak} W. R. Gibbs, Li Ai and W. B. Kaufmann, \pr C 57, 784(1998)

\bibitem{kh} W. B. Kaufmann and G. E. Hite, \prc 60, 055204(1999)

\bibitem{olsson} M. G. Olsson and W. B. Kaufmann,  Pion-nucleon Newsletter 
No. 16 (2002)

\bibitem{pavan} M. M. Pavan, R. A. Arndt, I. I. Strakovsky and R. L.
Workman, Pion-nucleon Newsletter No. 16 (2002)

\bibitem{gwu}  VPI/GWU phase-shift analysis, available on the Internet 
at \verb+http://gwdac.phys.gwu.edu/+

\bibitem{bkm} V. Bernard, N. Kaiser and U.-G. Meissner, Z. Phys. C 60,
111(1993)

\bibitem{gls} J. Gasser, H. Leutwyler and M. E. Sainio, \pl B 253, 260(1991)

\bibitem{bl} T. Becher and H. Leutwyler, Eur. Phys. J. C9, 643(1999)

\bibitem{brown} L. Brown, W. J. Pardee, and R. D. Peccei, Phys. Rev.
 {\bf D4}, 2801(1971)

\bibitem{sidmike} S. A. Coon and M. D. Scadron, J. Phys. G: Nucl. Part.
Phys. 18, 1923(1992)

\bibitem{dong} S. J. Dong, J.-F Laga\"e and K. F. Liu, \prd 54, 5496(1996)

\bibitem{chengli} Ta-Pei Cheng and Ling-Fong Li, ``Gauge Theory of
Elementary Particle Physics'', Clarendon Press 1984

\bibitem{cohen} T. D. Cohen, R. J. Furnstahl and D. K Griegel,  \prc 
45,1881(1992); \prl 67, 961(1991)

\bibitem{gammal} A. Gammal and T. Frederico, \prc 57, 2830(1998)

\bibitem{magda} M. Ericson, \np A577, 147c(1994)

\bibitem{ballot} J-L. Ballot, M. Ericson and M. R. Robilotta, 
\prc 61, 5202(2000)

\bibitem{gor} M. Gell-Mann, R. J. Oakes and B. Renner, \pr 175, 2195(1968) 

\bibitem{sigmod} V. Koch, Int. Jour. Mod. Phys. E Nucl. Phys. 6, 203(1997);
M. Gell-Mann and M. L\'evy, Nuovo Cimento 16, 705(1960)

\bibitem{gasser} J. Gasser, Pion-nucleon workshop 1984, p. 266
Los Alamos Report LA-11184-C, Ann. Phys. 136, 62(1981);
T. P. Cheng, \prd 13, 2161(1976)

\bibitem{gasserbis} J. Gasser 1981 op. cit. See eq. 12.3

\bibitem{jameson} I. Jameson, A. A. Rawlinson and A. W. Thomas, Aust.
J. Phys. 47, 45(1994)

\bibitem{coon} S. A. Coon and M. D. Scadron, \pr {\bf C42},
 2256(1990); \pr {\bf C23}, 1150(1981)

\bibitem{lissia} M. Lissia, M.-C. Chu, J. W. Negele and J. M. Grandy,
\np A 555, 272(1993)

\bibitem{gogohiatoki} V. Gogohia and H. Toki, \pr D 61, 036006(2000)

\bibitem{shuryak1} E. V. Shuryak, \pl 79B, 135(1978) 

\bibitem{gogohia} V. Gogohia and Gy. Kluge, \pr D 62, 076008(2000)

\bibitem{chodos1} A. Chodos, R. L. Jaffe, K. Johnson, C. B. Thorn and
V. F. Weisskopf, \prd 10, 2599(1974)

\bibitem{chodos2} A.  Chodos and Charles B. Thorn, \prd 12, 2733(1975)

\bibitem{barnhill}M. V. Barnhill, W. K. Cheng and A. Halprin, \prd 20,
727(1979) 

\bibitem{degrand} T. DeGrand, R. L. Jaffe, K. Johnson and J. Kiskis,
\prd 12, 2060(1975) 

\bibitem{chan} M. S. Chanowitz and S. Sharpe, \np B 222, 211(1983)

\bibitem{vasconcellos} C. A. Z. Vasconcellos, H. T. Coelho, F. G. Pilotto,
B. E. J. Bodmann, M. Dillig and M. Razeira, Eur. Phys. J. C 4, 115(1998)

\bibitem{birse} M. C. Birse and J. A. McGovern, \pl B 292, 242(1992)

\bibitem{jaffesig} R. L. Jaffe, \pr D 21, 3215(1980)

\bibitem{jameson1} I. Jameson, A. W. Thomas and G. Chanfray,
J. Phys. G: Nucl. Par. Phys. 18, L159(1992) 


\end{thebibliography}
\end{document}